\documentclass[11pt,a4paper]{article}
\pdfoutput=1
\usepackage[T1]{fontenc}
\usepackage[utf8]{inputenc} 
\usepackage{lmodern}

\usepackage{authblk}
\usepackage{graphicx}
\usepackage{amsmath, amssymb}
\usepackage{subcaption}
\usepackage{verbatim}
\usepackage{indentfirst}

\title{Study of the energy calibration of the DEAP-3600 detector using Na-22 source data and simulations}

\author[1]{L. Luzzi\thanks{on behalf of the DEAP-3600 collaboration}}
\affil[1]{Centro de Investigaciones Energéticas, Medioambientales y Tecnológicas,\\
Avenida Complutense 40, 28040, Madrid, Spain}
\affil[ ]{\texttt{ludovico.luzzi@ciemat.es}}

\date{} 

\begin{document}

\maketitle

\begin{abstract}
DEAP-3600 is a single-phase liquid argon (LAr) direct-detection dark matter experiment, operating 2 km underground at SNOLAB (Sudbury, Canada). The detector consists of 3.3 tons of LAr contained in a spherical acrylic vessel. At WIMP mass of 100 GeV, DEAP-3600 has a projected sensitivity of 10$^{-46}$ cm$^{2}$ for the spin independent elastic scattering cross section of WIMPs. Radioactive sources have been used for the energy calibration and to test the detector performance. One of the most effective calibration run has been taken with a $^{22}$Na source deployed in a tube located around the DEAP-3600 steel shell. The simultaneous emission of three $\gamma$s by the source provides an excellent tagging for the $^{22}$Na decay. The results concerning the energy response of the detector and the agreement between data and Monte Carlo simulations in DEAP-3600  are investigated in this study.
\end{abstract}



\section{Introduction}

DEAP-3600 (Dark matter Experiment using Argon Pulse-shape discrimination) \cite{a} is an  experiment for the direct search of Weakly Interacting Massive Particles (WIMPs). DEAP-3600 is located 2 km underground at SNOLAB in Sudbury, Canada. This detector features a target mass of (3269 $\pm$ 24) kg of liquid argon (LAr) and stands as the largest of its kind in operation \cite{b}. Cooldown for the third fill of DEAP-3600 will start in summer 2024. 

Given the impossibility of introducing radioactive sources within the active volume to prevent any contamination of the liquid argon target material, the study of the energy calibration of the detector is a particularly challenging aspect of the experiment. Thus, various radioactive sources in dedicated setups placed outside the detector have been employed for this purpose. 

$^{22}$Na is an artificial isotope with a half-life of 2.6 years. It decays to an excited state of $^{22}$Ne by $\beta$$^{+}$ (90.2\% branching ratio) or electron capture (9.7\%).  The emitted positron annihilates with an electron from the surrounding material, resulting in a characteristic emission at 511 keV. Due to momentum conservation, two $\gamma$s are produced and emitted in nearly opposite directions. After a few ps from the decay, the excited $^{22}$Ne state de-excites emitting a 1.27 MeV $\gamma$. Only a very small fraction (0.1\%) of the decays leads directly to the ground state of neon via $\beta$$^{+}$.

This topology, typically emitting three $\gamma$s almost simultaneously, offers a highly effective tagging method for identifying the interaction produced by the $^{22}$Na decay. Thus, this source serves as a powerful tool to assess the energy calibration of the detector and study the comparison of data and Monte Carlo (MC) simulations.

\section{Calibration setup}

The source, provided by the Eckert and Ziegler company, is housed in an A3015 capsule. This capsule is a 3 mm radius and 8 mm height stainless-steel cylinder, containing the active element. To comply with SNOLAB safety regulations, the source was wrapped in copper foil and sealed with a copper closure, measuring 20 mm in diameter and 9 mm in height. 

\begin{figure} [h]
    \centering
    \includegraphics[width=0.5\textwidth]{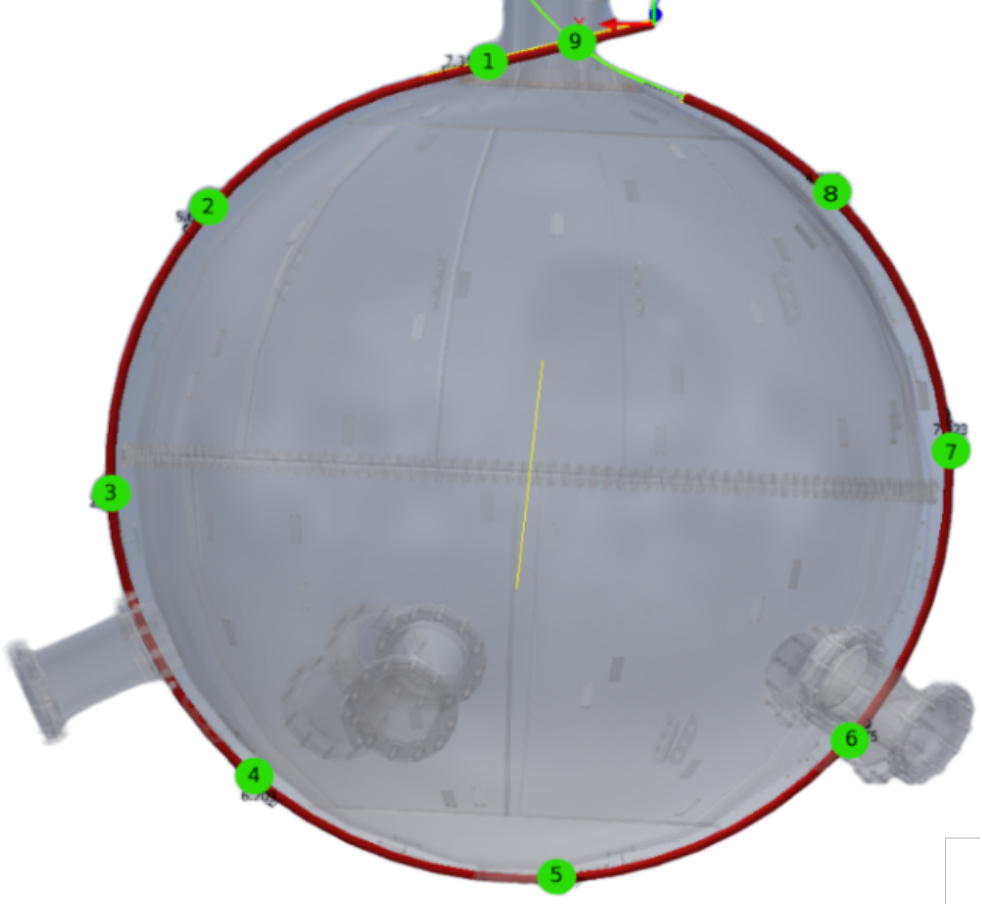}
    \caption{The CAL F calibration tube outside the outer steel shell in DEAP-3600.}
    \label{fig:cal_f}
\end{figure}

Adjacent to the copper enclosure on both sides, two LYSO crystals are placed. These high-density scintillator crystals are employed to detect $\gamma$ rays originating from the  $^{22}$Na decay. The crystals are wrapped with PTFE tape and connected to PMTs, which record the scintillation light emitted by the LYSO crystals. The entire setup is enclosed within a stainless-steel canister and attached to a deployment cable, enabling it to be positioned at various points inside an external calibration tube called CAL F \cite{a}. Nine pre-determined source positions were designated along the cable (Figure \ref{fig:cal_f}).
These calibration data were taken between November 2016 and April 2017.

\section{Photoelectron reconstruction}

In DEAP-3600 there are two methods to count the number of scintillation photons in a PMT pulse. 
A standard one consists in denoting the number of photoelectrons (PE) from dividing the area under the peak by the mean of the single-photoelectron (SPE) charge distribution: the result is a biased estimator, where the bias is caused mainly by the afterpulsing (AP) in the PMT. 

A less biased method consists in estimating the number of PE using a likelihood analysis: it assumes that the number of PE in a pulse is the contribution of PE from scintillation photons ($n_{Sc}$) and AP signals ($n_{AP}$). Given the charge of the pulse, the LAr probability density function (PDF), the times of the previous pulses, the time of the AP, the charge PDF and the SPE charge distribution, Bayes' theorem is applied and we can calculate how likely it is to observe a number of PE at the time of a given pulse. The values of $n_{Sc}$ and $n_{AP}$ are determined using the minimum mean square error estimator and then only $n_{Sc}$ is taken into account, removing in this way the AP contribution.

This second method is the one used in this report as the PE estimator. For more details refer to \cite{c}.

\section{Tagging algorithm}

LYSO is an excellent material to detect the $\gamma$ interaction due to the high light yield (32 photons/keV) and short attenuation length of 1.2 cm of the crystal for 511 keV $\gamma$s. 
The tagging technique relies on the signals of the PMTs placed on the side of the LYSO and on the coincidence between the events in the crystals and in LAr. 
The DAQ trigger is similar to the one of the physics run and consists in the coincidence of the PMTs inside the detector: to trigger the detector, a charge approximately equal to at least 19 PE must be detected across all 177 ns sliding window. Upon triggering, data from all PMTs are digitized from $-$2.6 to $+$14.4 $\mu$s relative to the trigger time \cite{c}.
Low-level cuts were applied to the data to reject noisy events: events with no charge information, pile up events, pre-trigger pile up and post-trigger pile up events, events with significant amount of early light pulses in the first 1600 ns of the waveform.

\begin{figure} [h]
    \centering
    \includegraphics[width=0.7\textwidth]{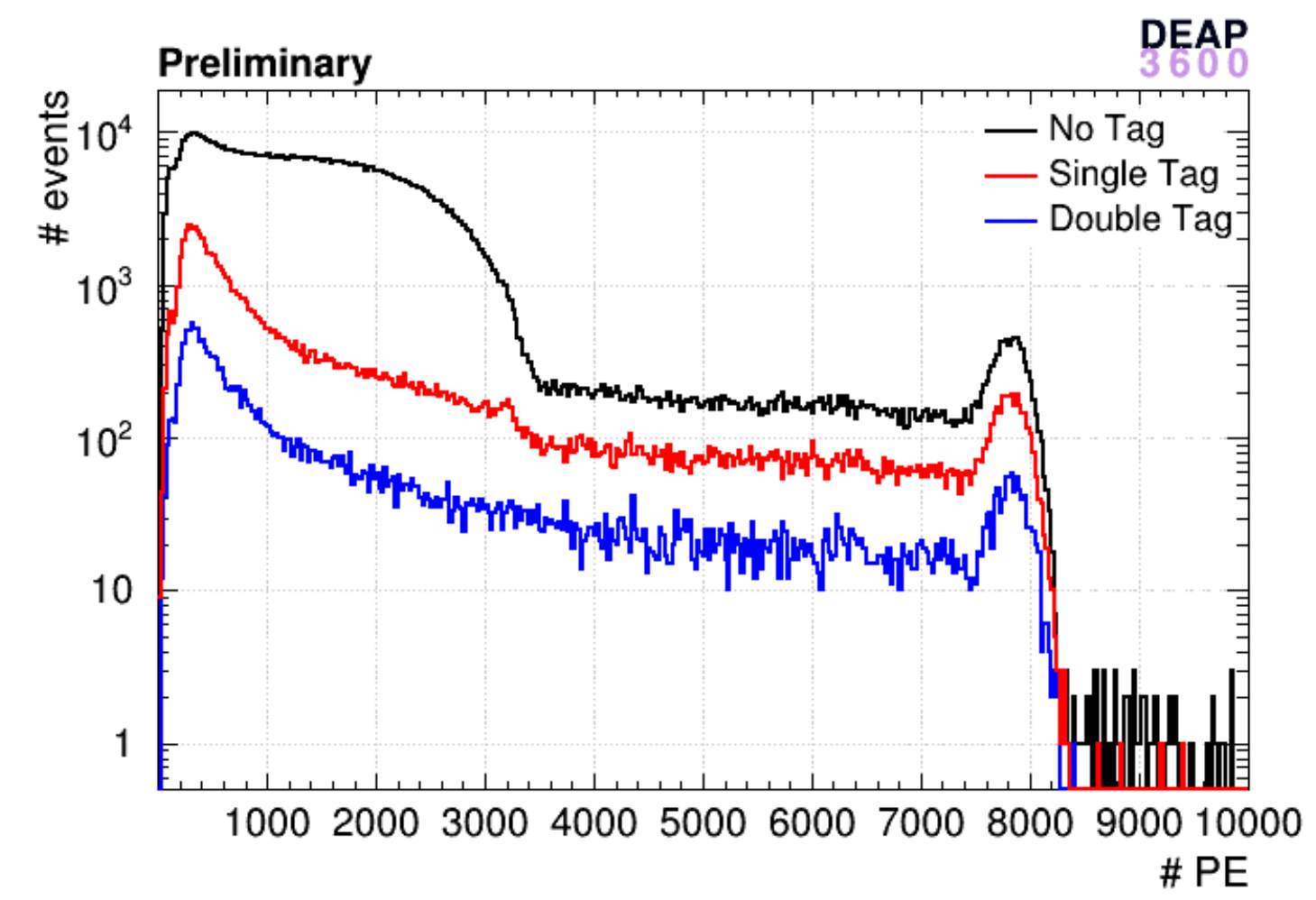}
    \caption{Energy spectrum in LAr for different tagging configurations for position 1.}
    \label{fig:spectrum}
\end{figure}

The energy spectrum of the $\gamma$s from the $^{22}$Na source, interacting with the LAr volume, shows variations under three different tagging conditions (Figure \ref{fig:spectrum}):

\begin{enumerate}
    \item no signal from the tagging PMTs: for energies <\textasciitilde{}3500 PE the uncorrelated events and untagged $^{22}$Na interactions are dominated by the $^{39}$Ar beta decay spectrum; 
    \item signal from just one of the tagging PMTs: in some cases this occurs when the 1.27 MeV $\gamma$ is detected by one of the tagging PMTs, and one of the 511 keV $\gamma$s reaches the LAr volume while the other 511 keV $\gamma$ is directed outside the detector; 
    \item double tagging: this represents the most stringent condition, where both 511 keV $\gamma$s are tagged by the tagging PMTs, and the 1.27 MeV interacts with the LAr.
\end{enumerate}

The full energy peak of the 1.27 MeV $\gamma$ is visible for all the tagging conditions while the 511 keV peak is only visible for the single tag configuration.

\section{Comparison between data and simulations}

Monte Carlo simulations of the $^{22}$Na source were performed using RAT, the official analysis tool of DEAP-3600, with the goal of comparing them with the data and understanding the performance of the detector. 
The simulation uses samples from RAT simulations which include the full optical model.
The energy response function is determined using the $^{39}$Ar beta decay spectrum, and $\gamma$s from the $^{22}$Na source and from isotopes present in detector materials ($^{40}$K, $^{214}$Bi, $^{208}$Tl). 

\begin{figure}[h]
    \centering

    \begin{subfigure}[b]{0.49\textwidth}
        \includegraphics[width=\textwidth]{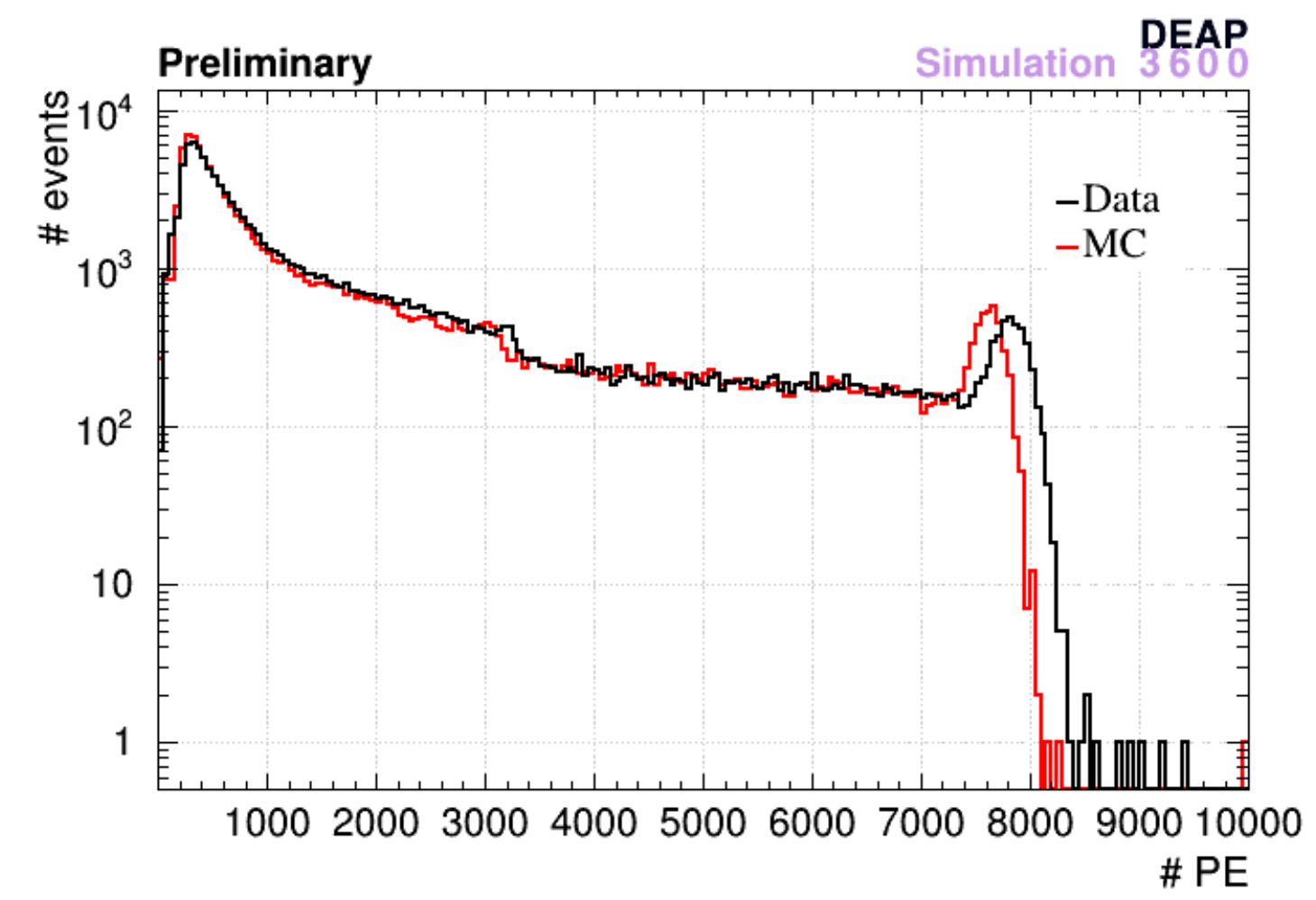}
        \caption{}
        \label{fig:pos_1_no_norm}
    \end{subfigure}
    \hfill
    \begin{subfigure}[b]{0.49\textwidth}
        \includegraphics[width=\textwidth]{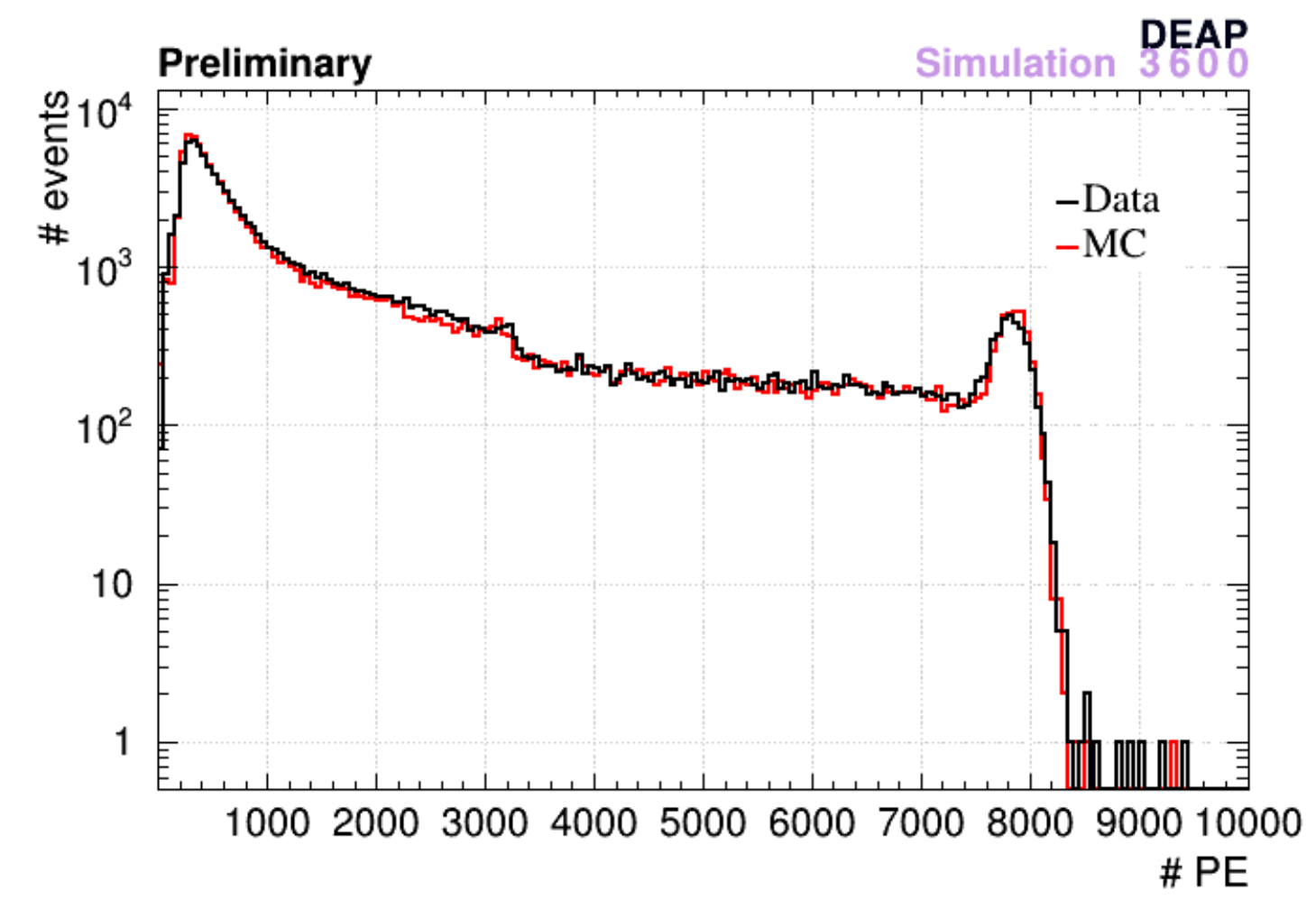}
        \caption{}
        \label{fig:pos_1_fit}
    \end{subfigure}


    \caption{Data (black) and MC (red) energy spectra for single tag events for position 1: Figure (a) shows the spectra before any correction applied; Figure (b) shows the spectra after the correction.}
    \label{fig:pos_1_spectra}
\end{figure}

Superimposed on the Figure \ref{fig:pos_1_no_norm} are the data and the MC spectra, for events of single tagging, for position 1: it can be seen that the shapes of the two spectra are similar but there is a shift in energy. By applying a Gaussian fit to the full energy peak of the two spectra, their position is determined. Based on this, a multiplicative correction factor of 1.03 is applied to the simulated events in order to match the data (Figure \ref{fig:pos_1_fit}). 
The correction factor has been measured for all the positions and it varies between 1.01 and 1.03.
MC predicts \textasciitilde{}1\% of asymmetry in light yield from top to bottom of the detector and \textasciitilde{}3\% of radial dependence of the response. Further investigation into the causes of this shift is ongoing.

\section{Energy Calibration}
\begin{figure}[h]
    \centering

    \begin{subfigure}[b]{0.49\textwidth}
        \includegraphics[width=\textwidth]{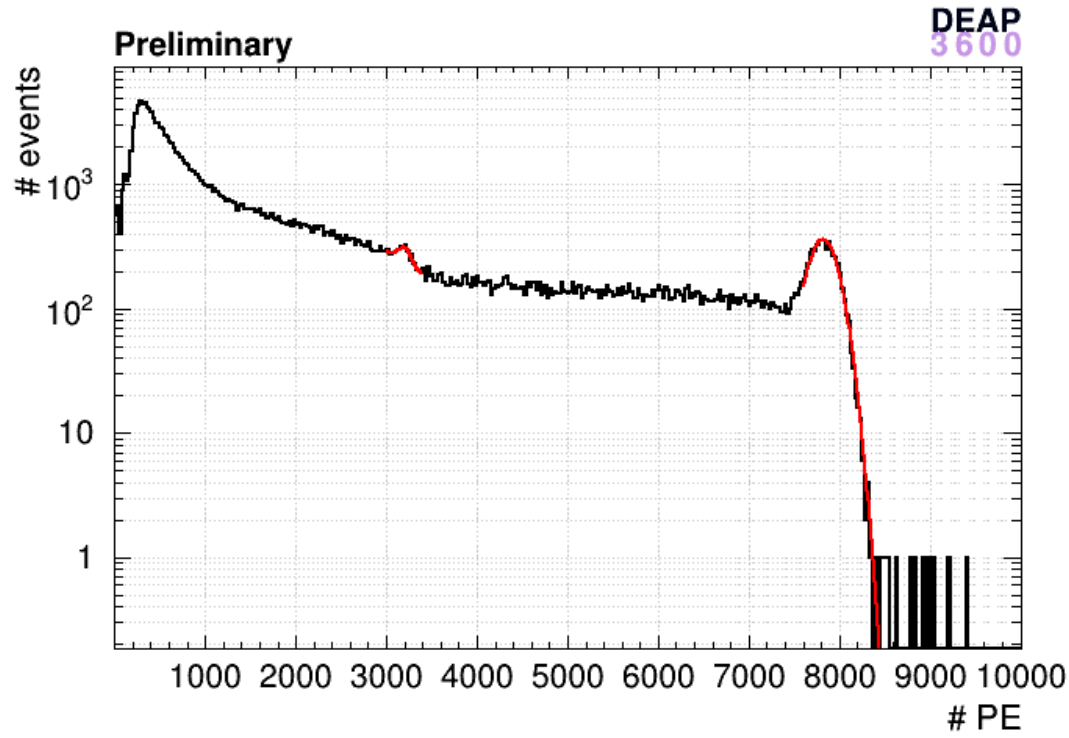}
        \caption{}
        \label{fig:spectrum_fit_full}
    \end{subfigure}
    \hfill
    \begin{subfigure}[b]{0.49\textwidth}
        \includegraphics[width=\textwidth]{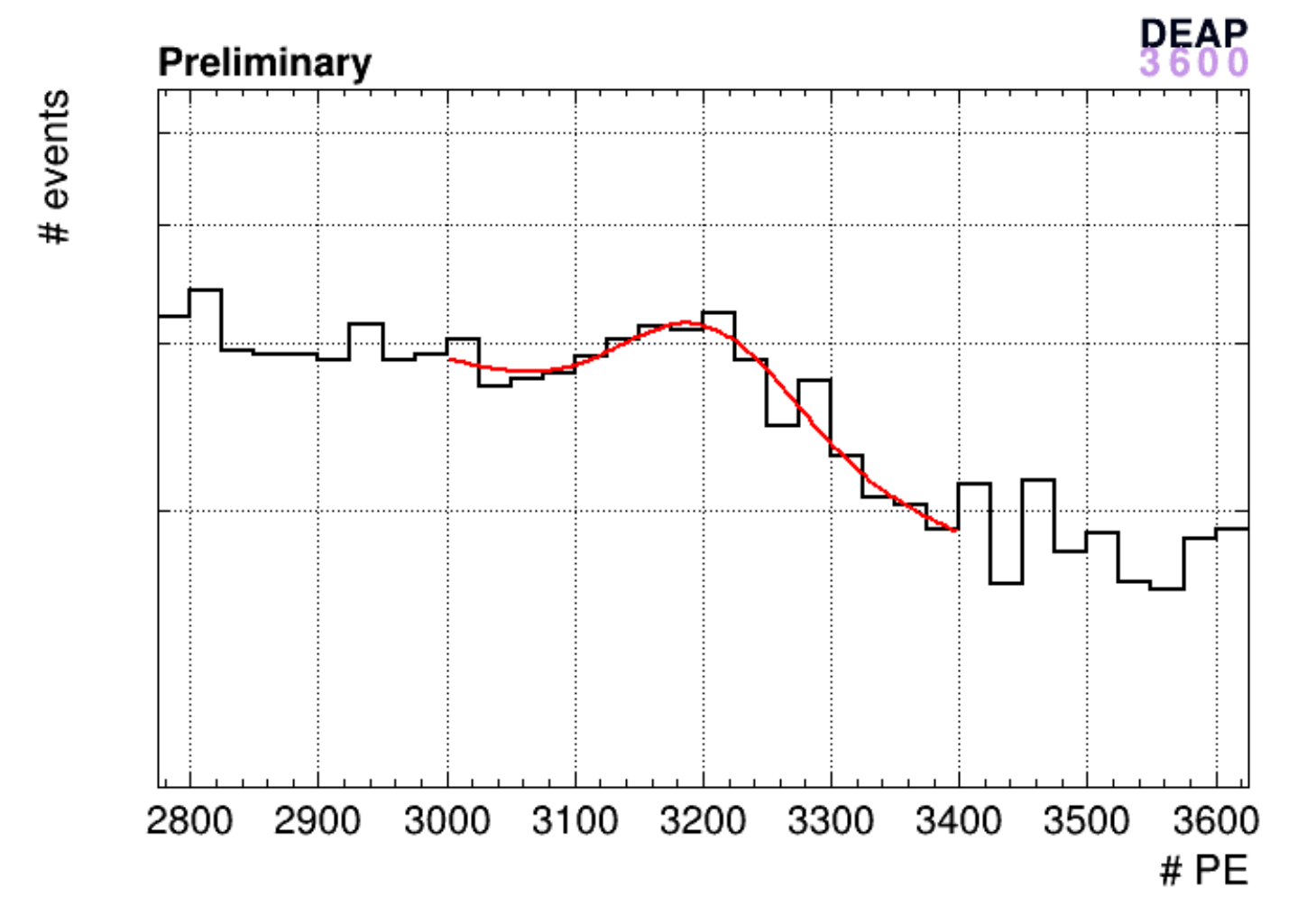}
        \caption{}
        \label{fig:spectrum_fit_511}
    \end{subfigure}


    \caption{On the left: fits performed on the 511 keV and 1.27 MeV peaks; on the right: a zoom of the fit on 511 keV peak.}
    \label{fig:spectrum_fit}
\end{figure}

Energy calibration can be performed using the $^{22}$Na spectrum. The analysis focuses on events with only one tag in the tagging PMTs, ensuring the visibility of the peak at 511 keV. A Gaussian + linear fit is performed for the 511 keV peak, and a Gaussian fit is performed for the 1.27 MeV peak (Figure~\ref{fig:spectrum_fit_full}). The mean values of the peaks resulting from the fit are then divided by the energy of those peaks in keV.
For a specific position, the weighted average of the light yield evaluated by the two peaks is performed, and this process is repeated for each position. Table \ref{tab:i} presents the calibration results obtained for each of the 9 source positions.

The difference between the light yields measured in the 9 positions is small.
The average light yield measured is 6.15 $\pm$ 0.02 PE/keV and it is in agreement with the measurements performed on the 511 keV peak and on the 1.27 MeV peak individually.

\begin{table}[htbp]
\centering
\caption{\label{tab:i} Energy calibration for each position of the source inside the CAL F tube.}
\smallskip
\begin{tabular}{|c|c|}
\hline
Position&Average LY (PE/keV)\\
\hline
1 & 6.13 $\pm$ 0.02 \\
2 & 6.13 $\pm$ 0.03 \\
3 & 6.19 $\pm$ 0.02 \\
4 & 6.16 $\pm$ 0.02 \\
5 & 6.20 $\pm$ 0.03 \\
6 & 6.19 $\pm$ 0.02 \\
7 & 6.17 $\pm$ 0.02 \\
8 & 6.11 $\pm$ 0.02 \\
9 & 6.10 $\pm$ 0.02 \\

\hline
\end{tabular}
\end{table}

\section{Conclusions}

Studying the energy calibration of the DEAP-3600 detector is a key aspect in understanding the performance of the experiment.
Data have been taken with the $^{22}$Na source encapsulated between two plastic scintillators and two PMTs and positioned around the outer steel shell of the detector in a calibration tube, at 9 pre-set positions. 
To reconstruct the number of PE in a PMT signal an unbiased PE estimator with after pulsing removed has been used.
The energy spectrum of the $\gamma$s from the $^{22}$Na source, interacting with the LAr volume, shows variations under three different tagging conditions. 
To improve the understanding of the detector, Monte Carlo simulations of the source were carried out at the same real positions. 
By selecting only single tagged events, data and MC spectra are compared, showing a similar shape but shifted in energy. Applying a correction factor to the MC spectrum it is possible to achieve the match between the spectra. This shows how the MC is able to reproduce data with good accuracy.
The energy calibration of the detector is also measured for each of the 9 positions, resulting in an average of 6.15 $\pm$ 0.02 PE/keV.

\clearpage

\section*{Acknowledgments}
We thank the Natural Sciences and Engineering Research Council of Canada (NSERC),
the Canada Foundation for Innovation (CFI),
the Ontario Ministry of Research and Innovation (MRI), 
and Alberta Advanced Education and Technology (ASRIP),
the University of Alberta,
Carleton University, 
Queen's University,
the Canada First Research Excellence Fund through the Arthur B.~McDonald Canadian Astroparticle Physics Research Institute,
Consejo Nacional de Ciencia y Tecnolog\'ia Project No. CONACYT CB-2017-2018/A1-S-8960, 
DGAPA UNAM Grants No. PAPIIT IN108020 and IN105923, 
and Fundaci\'on Marcos Moshinsky,
the European Research Council Project (ERC StG 279980),
the UK Science and Technology Facilities Council (STFC) (ST/K002570/1 and ST/R002908/1),
the Leverhulme Trust (ECF-20130496),
the Russian Science Foundation (Grant No. 21-72-10065),
the Spanish Ministry of Science and Innovation (PID2019-109374GB-I00) and the Community of Madrid (2018-T2/ TIC-10494), 
the International Research Agenda Programme AstroCeNT (MAB/2018/7)
funded by the Foundation for Polish Science (FNP) from the European Regional Development Fund,
and the European Union's Horizon 2020 research and innovation program under grant agreement No 952480 (DarkWave).
Studentship support from
the Rutherford Appleton Laboratory Particle Physics Division,
STFC and SEPNet PhD is acknowledged.
We thank SNOLAB and its staff for support through underground space, logistical, and technical services.
SNOLAB operations are supported by the CFI
and Province of Ontario MRI,
with underground access provided by Vale at the Creighton mine site.
We thank Vale for their continuing support, including the work of shipping the acrylic vessel underground.
We gratefully acknowledge the support of the Digital Research Alliance of Canada,
Calcul Qu\'ebec,
the Centre for Advanced Computing at Queen's University,
and the Computational Centre for Particle and Astrophysics (C2PAP) at the Leibniz Supercomputer Centre (LRZ)
for providing the computing resources required to undertake this work.


\clearpage

\begin{thebibliography}{99}

\bibitem{a}
DEAP-3600 Collaboration, \emph{{Design and construction of the DEAP-3600 dark matter detector}}, {{Astroparticle Physics}},  {{Volume 108, March 2019, Pages 1-23}}.

\bibitem{b}
DEAP-3600 Collaboration, \emph{{Precision measurement of the specific activity of $^{39}$Ar in atmospheric argon with the DEAP-3600 detector}}, {{The European Physics Journal C}},  {{ 83, 642 (2023)}}.

\bibitem{c}
DEAP-3600 Collaboration, \emph{{Pulse-shape discrimination against low-energy Ar-39 beta decays in liquid argon with 4.5 tonne-years of DEAP-3600 data}}, {{The European Physical Journal C}},  {{ 81, 823 (2021)}}.













\end{thebibliography}
\end{document}